\documentstyle[11pt]{article}
\font \msb=msbm10 scaled \magstep1
\newcommand{\bR}{\mbox{\msb R} }
\newcommand{\bC}{\mbox{\msb C} }

\font \eul=eufm10 scaled \magstep2

\newcommand{\gal}{\mbox{\eul g}}
\newcommand{\aal}{\mbox{\eul a}}
\newcommand{\rrq}{\mbox{\eul r}}

\def\l{{\lambda}}
\def\m{{\mu}}
\def\a{{\alpha}}
\def\b{{\beta}}
\def\g{{\gamma}}

\def\D{{\Delta}}
\def\bD{{\bf \Delta}}
\def\s{{\sigma}}

\def\ve{{\varepsilon}}
\def\P{{\Psi}}
\def\SL{{\cal L}}
\def\N{{\cal V}}
\def\tr{\mathop{\rm tr}}
\def\te#1{{\widetilde{#1}}}

\def\on#1#2{\mathop{\vbox{\ialign{##\crcr\noalign{\kern2pt}
$\scriptstyle{#2}$\crcr\noalign{\kern2pt\nointerlineskip}
\kern-2pt$\hfil\displaystyle{#1}\hfil$\crcr}}}\limits}
\def\nn{ \nonumber }
\def\bq{ \begin{equation} }
\def\eq{ \end{equation} }
\def\ben{ \begin{eqnarray} }
\def\en{ \end{eqnarray} }
\def\ll{ \label }
\def\frac#1#2{{#1\over #2}}
\def\dfrac#1#2{{\displaystyle{#1\over#2}}}
\def\dpr#1#2{{\displaystyle{{\partial #1}\over{\partial#2}}}}

\begin{document}
\title{Automorphisms of $sl(2)$ and
dynamical $r$-matrices.}
\author{
 A.V. Tsiganov\\
{\small\it
 Department of Mathematical and Computational Physics,
 Institute of Physics,}\\
{\small\it
St.Petersburg University,
198 904,  St.Petersburg,  Russia}
}
\date{}
\maketitle

\begin{abstract}
Two  outer automorphisms of infinite-dimensional
representations of $sl(2)$ algebra are considered.  The similar
constructions for the loop algebras and yangians are presented.
The corresponding linear and quadratic $R$-brackets include the
dynamical $r$-matrices.
\end{abstract}
\section{Introduction}
We are concerned with the representation theory of loop
algebras and yangians and with the method of separation of
variables in framework of the inverse scattering method.  Our
aim is to apply the representation theory of $sl(2)$ and
$sl(N)$ algebras into the $r$-matrix formalism.

Let us start with the following Levi-Civita theorem \cite{lc04}:  if
the Hamilton-Jacobi equation associated with hamiltonian
\bq
H=g^{ij}(x_1,\ldots,x_n)p_ip_j+h^j(x_1,\ldots,x_n)p_j+
U(x_1,\ldots,x_n)\,, \qquad \{p_i,x_j\}=\delta_{ij}\,,
\ll{gh}
\eq
can be integrated by separation of variables then the equation is
integrated with $U=0$ (\ref{gh}), i.e.  in the absence of a
force, in the same system of coordinates.

This note is devoted to the similar problem. Consider a classical
hamiltonian system completely integrable on a $2n$-dimensional
symplectic manifold $D$. It means that system possesses $n$
independent integrals $I_1,\ldots,I_n$ in the involution
\[\{I_i(x,p),I_j(x,p)\}=0\,,\qquad i,j=1,\ldots,n\,,\]
where $\{x_j,p_j\}_{j=1}^n$ is some coordinate system in $D$.

Introduce the following mapping
\bq
I_j\to I'_j=\sum_i^n a_{ij}(x,p)\cdot[I_i+v_i(x,p)]\,,\ll{mapi}
\eq
here $a_{ij}$ and $v_j$ are certain functions on $D$.

Under the suitable conditions for functions
$a_{ij}$ and $v_j$, the mapping (\ref{mapi}) could define a new
integrable system on $D$ with independent integrals
$I'_1,\ldots,I'_n$ in the involution
\[\{I'_i(x,p),I'_j(x,p)\}=0\,,\qquad i,j=1,\ldots,n\,,\] and could
preserve the property of separability in the same coordinate system.

Remind, that systems discovered by St{\"a}ckel (see review
\cite{pe91}) have the similar properties. Their integrals are
\bq
I_j=\sum_{i=1}^n a_{ij}(x_1,\ldots,x_n)[p_i^2+v_i(x_i)]\,,\ll{stint}
\eq
with the special functions $a_{ij}(x_1,\ldots,x_n)$ and the
arbitrary functions $v_i(x_i)$.

We try to make a first step to understand of the algebraic
roots of constraints on the functions $a_{ij}$ and $v_j$, which
guarantee the properties of integrability and separability for
mapping (\ref{mapi}) in the framework of the inverse scattering
method.


\section{Outer automorphisms of representations of $sl(2)$}
Let $W$ be an infinite-dimensional representation of the Lie algebra
$sl(2)$ in linear space $V$ defined in the Cartan-Weil basis
$\{s_3,s_\pm\}\in \rm{End}(V)$ equipped with the natural bracket
\bq [s_3,s_\pm]=\pm s_\pm\,,\qquad
[s_+,s_-]=2s_3
\eq
and the single Casimir operator
\bq
\bD=s_3^2+\dfrac12(s_+s_-+s_-s_+)\,.
\eq
If operator $s_+$ is invertible in $\rm{End}(V)$ then the mapping
\ben
&&s_3\to s_3'=s_3\,,\qquad s_+\to s_+'=s_+\,, \nn\\
&&s_-\to s_-'=s_-+fs_+^{-1}\,,\qquad f\in\bC \ll{mapp}
\en
is an outer automorphism of the space of
infinite-dimensional representations of $sl(2)$ in
$V$. The mapping (\ref{mapp}) shifts the spectrum of $\bD$ on
the parameter $f$
\bq \D \to \D'=\D+f. \eq
Let us call the mapping (\ref{mapp}) an additive automorphism.

Assuming in addition the value of Casimir operator $\bD$ is
equal to zero $\D=0$ in $W$, then the mapping
\ben
&&s_j\to s_j'=s_j\cdot(1-fs_+^{-1})\,,\qquad j=\pm,3\,,\qquad
f\in\bC\nn\\
\ll{mapmu}\\
&&\bD\to \bD'=\bD\cdot(1-fs_+^{-1})^2\,,\qquad\D=\D'=0\,, \nn
\en
is another outer automorphism of representation $W$.  Let us
call the mapping (\ref{mapmu}) a multiplicative automorphism.

These special infinite-dimensional representations $W$ could be
obtained by using the previous additive automorphism
(\ref{mapp}).  Notice, that the similar representations with
$\D=0$ are well known in the quantum conformal field theory
\cite{fk94}.

We can suppose that additive (\ref{mapp}) and multiplicative
(\ref{mapmu}) automorphisms define the one-parametric
realizations $W(f)$ of $sl(2)$.

For instance, realization of $sl(2)$ with one free parameter $f$ in
the classical mechanics  is given by
\bq s_3=\dfrac{xp}2\,,\quad
s_+=\dfrac{x^2}2\,,\quad s_-=-\dfrac{p^2}2+\dfrac{f}{x^2}\,,
\quad \D'=f\,,\ll{clrep}
\eq
where $(x,p)$ is a pair of canonical coordinate and momenta with
the classical Poisson bracket $\{p,x\}=1$. Similar realization of
generators $s_j$ as the differential operators in quantum case is
\bq
s_3=x\partial_x-l\,,\quad s_+=x\,,\quad
s_-=x\partial_x^2-2l\partial_x+\dfrac{f}{x}\,, \quad \D'=l(l+1)+f\,.
\nn \eq
For the quantum Calogero system
the more sophisticated representation of $sl(2)$ is equal to
\[s_3=\dfrac14\sum_{k=1}^N(x_kD_k+D_kx_k)\,,\quad
s_+=2\sum_{k=1}^Nx_k^2 \,,\quad
s_-=-2\sum_{k=1}^ND_k^2\,,\quad
\]
where $x_k$ are coordinates and $D_k$ are the corresponding Dunkl
operators \cite{dan89}.

Using the inner automorphisms of $sl(2)$ the one parametric mappings
(\ref{mapp}) and (\ref{mapmu}) can be generalized.  Let $W$ be an
infinite-dimensional representation of $sl(2)$ in $V$ defined
by operators $s_j\in {\rm End}\,(V)$ with the standard bracket and
the single quadratic Casimir operator $\bD$
\bq [s_i,s_j]=\ve_{ijk}s_k\,,\qquad\bD=s_1^2+s_2^2+s_3^2\,. \ll{pb}
\eq

Introduce the general linear operator $b\in {\rm End}\,(V)$ as
\bq b=\a_1s_1+\a_2s_2+\a_3s_3\,,\qquad \a_j\in\bC\,, \ll{bp} \eq
where $\a_j$ are three arbitrary parameters.
If $b$ is invertible operator, then the mapping
\bq s_j\to s'_j=s_j+\a_jb^{-1}
\ll{mapj}\eq
is an outer automorphism of the space of
infinite-dimensional representations of $sl(2)$ in
$V$, if the parameters $\a_j$ lie on the cone
\bq
\a_1^2+\a_2^2+\a_3^2=0\,.\ll{cone}
\eq
For this mapping the spectrum of the Casimir operator $\bD$
(\ref{pb}) is additively shifted on the parameter $f=\sum \a_j$
\bq
\D\to\D'=\D+f=\D+\sum_{j=1}^3\a_j\,.
\eq

If the value of Casimir operator $\bD$ is equal to zero $\D=0$
and the general linear operator $b\in {\rm End}\,(V)$ (\ref{bp}) is
invertible, then the mapping \bq s_j\to
s_j'=s_j(1-b^{-1})\,,\qquad\bD\to\bD'=\bD\cdot(1-b^{-1})^2\ll{mapmug} \eq
is an outer multiplicative automorphism of representation $W$.

Automorphisms (\ref{mapj}) and (\ref{mapmug}) define the general
two and three-parametric realizations $W(\a_j)$ of $sl(2)$.

For the $sl(N)$ algebra the similar additive automorphism shifted
the highest central element  of $N$-th order can be taken in
the form
\bq
s_{ij}\to s'_{ij}=s_{ij}+\dfrac{ A^{(m-1)}_{ij} }{ B^{(m)} }\,,
\qquad s_{ij}\in{\rm End}(V)\,,
\qquad m=N(N-1)/2\,.
\eq
In operators $s_{ij}$ functions $B^{(m)}$ and $A^{(m-1)}_{ij}$
are certain polynomials of degrees $m$ and $m-1$. Here we shall
not go into details of this problem, one example related to
$sl(3)$ will be presented in Section 4.

Motivated by realization (\ref{clrep}) we present one
application of automorphism (\ref{mapp}) in the theory of
integrable systems.  Consider a classical Hamiltonian system
completely integrable on phase space $D=\bR^{2n}$
with the natural hamiltonian
\[H=\sum_{j=1}^n p_j^2+V(x_1,\ldots,x_n)\,.\]
Let the phase space be identified completely or partially
with the $m$ coadjoint orbits in $sl(2)^*$ as (\ref{clrep}).
Then the mapping
\bq
H\to H'=H+\sum_{j=1}^m \dfrac{f_j}{x_j^2}\,,\qquad f_j\in\bR
\ll{pr1}
\eq
preserves the properties of integrability and separability.
The list of such systems can be found in \cite{pe91}.

The main our aim is in developing  similar constructions for
the loop algebras and yangians. Some examples of the integrable
systems related with this approach have been considered
in \cite{eekt94,ts94,ts95}.


\section{Dynamical $r$-matrices associated to $\te{sl(2)}$}
\setcounter{equation}{0}
All details of the general $r$-matrix scheme can be found in review
\cite{rs87} and in references therein.  We recall briefly
necessary elements
of $r$-matrix scheme for loop algebra $\gal=\te{sl(2)}$
to assume the standard identification of the dual spaces.

The loop algebra $\gal=\te{sl(2)}$ consists of Laurent
polynomials $s_j(\l)=\sum_k s_j\l^k\,,~j=1,2,3$ of spectral
parameter $\l$ with coefficients in $\aal=sl(2)$ and commutator
$[s_i\l^l,s_j\l^m]=\ve_{ijk}s_k\l^{l+m}$.  The standard
$R$-bracket associated with loop algebra $\gal$ is defined by
the following decomposition of $\gal$ into a linear sum of two
subalgebras
\bq \gal=\gal_+\on{+}{\,.}\gal_-\,,\qquad
\gal_+=\oplus_{i\geq 0}\aal\l^i\,,\qquad
\gal_-=\oplus_{i<0}\aal\l^i\,,\qquad
R=P_+-P_-\,.
\ll{dec}\\
\eq
Here $R$ is a standard $r$-matrix and
$P_\pm$ means the projection operators onto $\gal_\pm$ parallel
$\gal_\mp$.

The Lax equation may be presented in the form
\bq
\dfrac{dL(\l)}{dt}=-{\rm ad}^*_{\gal}A\cdot L\,,\qquad
A=\dfrac12 R(dP(L))\,,
\ll{lax}
\eq
where Lax matrix $L(\l)$ belongs to $\gal^*$, $P$ is an ${\rm
ad}^*$-invariant polynomial on $\aal^*$. For algebra $\te{sl(2)}$
polynomial $P(L)$ is a function of the unique invariant polynomial
$\D(\l)=\sum_{j=1}^3 s_j^2(\l)$.  Let $P(L)=\phi(\l)\D(\l)$, where
$\phi$ is a functions of $\l$. The integrals of motion $I_k$  related
to flow (\ref{lax}) are
\bq
I_k(L)={\rm Res}_{\l=0}(\phi_k(\l)\D(\l))\,, \ll{ham} \eq
where $\phi_k(\l)$ are various functions of spectral parameter
defining a complete set of integrals of motion \cite{rs87}.

The Lax matrix $L(\l)$ (\ref{lax}) is defined on the whole
infinite-dimensional phase space $\gal^*$. It is well known that the
standard $R$-bracket associated with (\ref{dec}) has also a large
collection of finite-dimensional Poisson ($ad^*_R$-invariant)
subspaces \cite{rs87}
\bq
{\SL}_{M,N}=\oplus^{N}_{j=-M}\,{\aal}^*\,\l^j\,, \qquad{\rm
provided}{~} M\geq 0\,;{~}N\geq 1\,.\label{subsp}
\eq
and, as a rule, the concrete physical systems are related to
the restrictions of the flow (\ref{lax}) to certain
low-dimensional Poisson submanifolds ${\SL}_{M,N}$
(\ref{subsp}).

The $r$-matrix scheme is extended easily to the twisted subalgebras
of loop algebra $\gal$ and corresponding $r$-matrices have rational,
trigonometric and elliptic dependence on spectral parameter.  We
shall work with a tensor form of the Lax equations and $R$-bracket on
$\te{sl(2)}$, that allows us to consider all the $r$-matrices
simultaneously. In addition, for brevity, we shall consider the
$\te{sl(2,\bR})$ only.

The Lax matrix $L(\l)$ (\ref{lax}) associated to the
hamiltonian (\ref{ham}) is given by
\bq
L(\l)=\sum_{k=1}^3 s_k(\l)\s_k\,,
\qquad \D(\l)=
\dfrac12 {\rm tr}L^2(\l)=-{\rm det} L(\l)\,,\ll{lmat}
\eq
where $\s_j$ are Pauli matrices. The $R$-bracket related to
decomposition (\ref{dec})  take the following form
\bq
\{{\on{L}{1}}(\l),{\on{L}{2}}(\m)\}=
[r_{12}(\l,\m),{\on{L}{1}}(\l)]-[r_{21}(\l,\m),{\on{L}{2}}(\m)\,]\,,
\ll{rpoi}
\eq
where the standard notations are introduced:
\ben
&&{\on{L}{1}}(\l)= L(\l)\otimes I\,,\qquad
{\on{L}{2}}(\m)=I\otimes L(\m)\,,\nn\\
\ll{r}\\
&&r_{12}(\l,\m)=\sum\limits^3_{k=1}w_k(\l,\m)\cdot
\sigma_k\otimes\sigma_k\,\qquad
r_{21}(\l,\m)=\Pi r_{12}(\m,\l)\Pi\,.\nn
\en
Here $\Pi$ is the permutation operator of auxiliary spaces and
$w_k(\l,\m)$ are  certain functions of spectral parameters only.
Their explicit dependence on $\l,\mu$ is not important for the moment.

Returning now to the additive automorphism of representations of $sl(2)$
(\ref{mapj}) we introduce the related mappings on the loop algebras
$\gal=\te{sl(2,\bR)}$ or $\gal=\oplus^n\te{sl(2,\bR)}$.

The entries of the Lax matrix are  Laurent polynomials with
coefficients in $sl(2)$. Let us consider the mapping
transforming these coefficients.  If parameters $a_j$ lie on
the cone (\ref{cone}), then the first mapping similar to the
automorphism (\ref{mapj})
\bq
s_j(\l)=\sum_k s_j\l^k\to s_j'(\l)=\sum_k [s_j+\a_jb^{-1}]\l^k\,,
\qquad b=\sum_{j=1}^3 \a_j s_j\,,
\ll{map1}
\eq
is a canonical change of variables on $sl(2)$.  This mapping is
defined for the infinite-dimensional representations of $sl(2)$ and
mapping (\ref{map1}) is a Poisson map with respect to the first
natural Lie-Poisson bracket and to the second linear Lie-Poisson
bracket associated with the standard $R$-matrix (\ref{dec}) on
$\te{sl(2)}$. For the certain Lax representation $L(\l)$
mapping (\ref{map1}) transforms the integrals of motion $I_k$
similar to ({\ref{pr1}}).

Let us introduce the general linear operator $b(\l)\in \te{sl(2)}$
\bq b(\l)=\sum_{j=1}^3 \a_j(\l) s_j(\l)\,,\ll{bl}
\eq
where $\a_j(\l)$ are functions of spectral parameter and of central
elements on $\te{sl(2)}$.  If functions $\a_j(\l)$ lie on the cone
\bq
\a_1^2(\l)+\a_2^2(\l)+\a_3^2(\l)=0\,,\ll{restr1}
\eq
then the second mapping similar to the automorphism
(\ref{mapj})
\bq
s_j(\l)\to s'_j(\l)=s_j(\l)+\a_j(\l)b^{-1}(\l)\,.\ll{map2}
\eq
is a Poisson map with respect to the first natural Lie-Poisson
bracket.  For the mapping (\ref{map2}) the Lax matrix
(\ref{lmat}), the invariant polynomial $\D(\l)$  and integrals of
motion $I_k$ (\ref{ham}) are additively shifted
\ben
&&L(\l)\to L'(\l)=L(\l)+\sum_{k=1}^3\a_k\s_k\cdot b^{-1}(\l)\,,\ll{shl}\\
\nn\\
&&\D(\l)\to \D'(\l)=\D(\l)+V(\l)=\D(\l)+\sum_{k=1}^3\a_k(\l)\,.\ll{det}\\
\nn\\
&&I_k\to I'_k=I_k+U_k\,,\qquad
U_k={\rm Res}_{\l=0}\left(\phi_k(\l)V(\l)\right)\in\bC\,.
\en
Restriction (\ref{restr1}) guarantees that initial integrals of
motion $I_k$ (\ref{ham}) transform by some constants
$U_k\in\bC$ (\ref{mapi}).  Hence, two Lax matrices $L(\l)$ and
$\L'(\l)$ (original and the image of mapping (\ref{map2}))
correspond to the same integrable system.

The entries of the initial matrix $L(\l)$ belong to $\gal^*$.  The
entries of the matrix $L'(\l)$ are the Laurent polynomials of
spectral parameter $\l$ with coefficients from the universal
enveloping algebra of $\aal^*$.  Nevertheless, the second Poisson
bracket $\{L'(\l),L'(\mu)\}$ can be directly calculated, because of
all the necessary Poisson brackets between $s_j(\l)$ and $b(\l)$ are
preassigned by (\ref{rpoi}).

Further we consider a special class of the linear dynamical
$R$-bracket related to $\te{sl(2)}$ and defined by the following
second restriction on coefficients $\a_j(\l)$
\bq \{ b(\l),b(\m)\}=g(\l,\m)b(\l)-g(\m,\l)b(\m)\,,
\qquad \l,\m\in\bC\,,\ll{restr2}
\eq
or, that is equivalent,
\bq
w_j(\l,\m)\a_j(\m)\a_i(\l)-
w_i(\l,\m)\a_i(\m)\a_j(\l)=
g(\l,\m)\cdot\a_k(\l)\,,
\label{rr2}
\eq
where $(j,i,k)$ are cyclic permutations of indices $(1,2,3)$ and
the scale function $g(\l,\m)$ depends of the spectral parameters
only.  This restriction is closely related with the separation of
variables method.  It guarantees that all zeroes of $b(\l)$ are
mutually commuting.  Below we assume that both conditions
(\ref{restr1}) and (\ref{restr2}) are always fulfilled for the
mapping (\ref{map2}).

In this case we get the Lax matrix $L'(\l)$
(\ref{shl}) obeys the linear $R$-bracket (\ref{rpoi}), where
constant $r_{ij}$-matrices substituted by $r_{ij}'$-matrices
depending on dynamical variables
\bq
r_{12}(\l,\mu)\to r'_{12}=r_{12}+\sum_{i,j=1}^3 \a_{ij}
(\l,\m)\,\sigma_i\otimes\sigma_j\,,\ll{rsb}\\
\eq
with coefficients $\a_{ij}$ being
\bq
\a_{ij}\,(\l,\m)= g(\l,\m)
\dfrac{\a_j(\m)\a_i(\m)w_i(\l,\m)}{b^2(\m)}
-
g(\m,\l)\dfrac{\a_i(\l)\a_j(\l)w_j(\l,\m)}{b^2(\l)}\,.
\eq
The proof see in \cite{krw95}.

Dynamical matrices $r'_{jk}(\l,\m)$ (\ref{rsb}) obey
the classical dynamical Yang-Baxter equation
\ben
&&[r'_{12}(\l,\m),r'_{13}(\l,\nu)]+
[r'_{12}(\l,\m),r'_{23}(\m,\nu)]+
[r'_{32}(\nu,\m),r'_{13}(\l,\nu)]+\nn\\
\ll{yb}\\
&&+[L_2'(\m),r'_{13}(\l,\nu)]-[L_3'(\nu),
r'_{12}(\l,\m)]+
[X_{123}(\l,\m,\nu),L_2'(\m)-L_3'(\nu)]=0\,,\nn
\en
where an explicit expression of the tensor $X(\l,\m,\nu)$ is
not important for the moment (see \cite{eekt94,krw95,hz95}).
This equation is the image of a standard $2$-cocycle in the new
object, which consists of Laurent polynomials of spectral
parameter $\l$ with coefficients from the corresponding
universal enveloping algebra.

We can suppose that mapping (\ref{map2}) defines family of Lax
matrices $L'(\l)$ and  dynamical $r$-matrices (\ref{rsb}) with
a fixed set of parameters $\a_j(\l)$ for certain integrable
system.  Thus, the Lax matrix $L(\l)$ (\ref{lax}), the constant
$r$-matrices (\ref{r}) and the standard classical Yang-Baxter
equation on $\te{sl(2)}$ are the limits of the dynamical ones
at $\a_j(\l)=0$.

The mapping (\ref{map2}) allows us to construct an
infinite set of the Lax matrices associated to loop algebra
$\te{sl(2)}$, which correspond to  different integrable systems.
According to \cite{krw95} we introduce an infinite set of mappings
\bq
s_j(\l)\to
s'_j(\l)=s_j(\l)+\left[\a_j(\l)b^{-1}(\l)\right]_{M_jN_j}\,,
\ll{mapmn}
\eq
where $[z]_{MN}$ means restriction of $z$ onto the certain Poisson
subspase $\SL_{MN}$ (\ref{subsp}) of the standard $R$-bracket
(\ref{dec}).

As an example, we can use the linear combinations of the
following Laurent and Fourier projections
\ben
{[ z ]_{MN}}&=&\left[\sum_{k=-\infty}^{+\infty} z_k\l^k\,
\right]_{MN}\equiv
\sum_{k=-M}^{N} z_k\l^k\,, \nn\\
&&\label{cutmn}\\
{[ z ]_{MN}}&=&\left[\sum_{k=-\infty}^{+\infty}
z_k\exp(k\cdot\l)\,\right]_{MN}\equiv
\sum_{k=-M}^{N} z_k\exp(k\cdot\l)\,,\nn
\en
if the corresponding $r$-matrices have rational or
trigonometric dependence on spectral parameter \cite{krw95}.

Generally speaking, we can not describe now all the possible
restrictions of the mapping (\ref{map2}) to certain
low-dimensional submanifolds, which allow us to define the second
Poisson bracket on $\te{sl(2)}$.  However, for a fairly large
class of properly defined projections (\ref{cutmn}) the mappings
(\ref{mapmn}) are the Poisson maps with respect to the second
$R$-bracket and the corresponding dynamical $r$-matrices obey the
dynamical Yang-Baxter equation (\ref{yb}).

For instance, an application of projections (\ref{cutmn}) yields
dynamical $r$-matrices
\bq r_{12}(\l,\mu)\to
r'_{12}=r_{12}+\sum_{i,j=1}^3 \a_{ij}
(\l,\m)\,\sigma_i\otimes\sigma_j\,,
\eq
with the following coefficients
\bq
\a_{ij}\,(\l,\m)= g(\l,\m)w_i(\l,\m)
\left[\dfrac{\a_j(\m)\a_i(\m)}{b^2(\m)}\right]_{MN}
-
g(\m,\l)w_j(\l,\m)
\left[\dfrac{\a_i(\l)\a_j(\l)}{b^2(\l)}\right]_{MN}.
\eq

The essential feature of the restricting mappings (\ref{mapmn})
is, in comparison with the mapping (\ref{map2}), that the
invariant polynomial $\D(\l)$ and all integrals of motion
$I_k$ are shifted now on the items depending on dynamical
variables
\ben
&&\D(\l)\to\D_{MN}(\l)=\D(\l)+V_{MN}(s_j,\a_j,\l)\,,\nn\\
\ll{nint}\\
&&I_k\to
I'_k=I_k+{\rm Res}_{\l=0}(\phi_k(\l) V_{MN}(s_j,\a_j,\l))\nn
\en
where $V_{MN}$ being
\bq V_{MN}(s_j,\a_j,\l)=\sum_{k=1}^3
\left(2s_k(\l)+[a_j(\l)b^{-1}(\l)]_{MN}\right)
[a_j(\l)b^{-1}(\l)]_{MN}\,.\ll{pot}
\eq
Hence, images of integrals of motion $I'_k$ of the
mapping (\ref{mapmn}) are functionally different from the original
ones $I_k$.

The second feature of the mapping (\ref{mapmn}) is that it
can be also applied to the finite-dimensional
representations of $sl(2)$. Consider the multipole
Lax matrices related to the rational $r$-matrix \cite{rs87}. Let
$\gal=\oplus^n sl(2,\bR)$ and elements of mapping (\ref{map2}) are
\bq
\a_1=1\,,\quad\a_2=i\,,\quad\a_3=0\,,\quad
b(\l)=\sum_{k=1}^n \dfrac{s_+^{(k)}}{\l-e_k}\,,
\quad e_j\neq e_k\in\bR\,.
\eq
The Taylor projection of $b^{-1}(\l)$
is determined by the following recurrence relations \cite{eekt94}
\bq
\left.\left[b^{-1}(\l)\right]_{MN}\right|_{M=0}
=\sum_{k=0}^N \N_k\l_{N-k}\,,\qquad
\N_k=\sum_{i=1}^n \left( s_+^{(i)}\sum_{j=0}^{k-1}
\N_{k-1-j}e_i^j\right)\,,
\qquad \N_0=1\,.\ll{tey}
\eq
The Taylor projection (\ref{tey}) is the well-defined
polynomial of the nilpotent operators $s_+^{(k)}$ without the
negative powers, which can be also used for finite-dimensional
representations of $sl(2)$.

Once again, the mappings (\ref{mapmn}) are not just
isomorphisms of the Lax matrices $L(\l)$ and $L'(\l)$, in
contrast with mapping (\ref{map2}), but they also preserve the
second Poisson $R$-bracket together with all the commuting
integrals of motion $I'_k$, i.e. preserve the properties of
integrability and separability.

Thus, the mappings (\ref{mapmn})
play the role of a dressing procedure allowing to construct the
Lax matrices $L'_{MN}(\l)$ for an infinite set of new
integrable systems starting from the single known Lax matrix
$L(\l)$ associated to one integrable model.

Now we shall briefly discuss the second
multiplicative automorphism (\ref{mapmug}). Let us start with some
Lax matrix $L(\l)\in \te{sl(2)}^*$. This matrix relates to the
integrable system with the Lax representation (\ref{lax})
and with the following integrals of motion (\ref{ham})
\[I_k={\rm Res}_{\l=0}\,(\phi_k(\l)\D(\l))\,.\]

Recall, that the multiplicative automorphism (\ref{mapmug}) of
$sl(2)$ transforms the Casimir operator by the rule
\[\bD\to\bD'=\bD\cdot\varphi(b)=\bD\cdot(1-b^{-1})^2\,.\]

Motivated by this transformation we consider the new
set of integrals $I'_k$ (see \ref{mapi}) defined by
\bq
I'_k={\rm Res}_{\l=0}\,(\varphi_k(b,\l)\cdot\D(\l))\,,\ll{intmul} \eq
where $\varphi_k(b,\l)$ are certain functions of $b(\l)$
(\ref{bl}) and of the spectral parameter $\l$.
The involution conditions
\[\{I_i',I_j'\}=0\,,\qquad i,j=1,\ldots,n\,,\]
yield the set of equations in $\varphi_k(b,\l)$
\[{\rm Res}_{\l\mu=0}\{\D(\l),b(\mu)\}\varphi_j(\l)\varphi_k(\mu)
\left(\dpr{\ln \varphi_j}{b}(\l)\D(\l)-
\dpr{\ln \varphi_k}{b}(\mu)\D(\mu)\right)=0\,.
\]
New integrals of motion $I_j'$ could be functionally different
from the original ones $I_j$.

In the quantum case the Poisson brackets should be replaced by
the standard commutator relations on $sl(2)$ and the
non-dynamical linear $R$-matrix bracket (\ref{rpoi}) becomes
the following commutator relations
\bq
\left[{\on{L}{1}}(\l),{\on{L}{2}}(\m)\right]=
[\rrq_{12}(\l,\m),{\on{L}{1}}(\l)]-[\rrq_{21}(\l,\m),{\on{L}{2}}(\m)\,]\,,
\quad \rrq(\l,\m)=-i\hbar r(\l,\m)\,.
\ll{rqb}
\eq
Theory of the general quantum linear $R$-bracket with dynamical
$r$-matrices is not  well developed  yet, but a nice feature of the
presented Poisson mappings (\ref{map2}) and (\ref{mapmn}) is that it
admits a natural quantization. All the necessary commutator
relations between operators $s_j(\l)$ and linear invertible operator
$b(\l)$ (\ref{bl}) are preassigned by (\ref{rqb}) and, therefore, the
introduction of the quantum dynamical $r$-matrices (\ref{rsb}) is a
straightforward calculation.  A similar calculation yields the direct
quantum counterpart of the dynamical Yang-Baxter equation.


\section{Dynamical $r$-matrices and separation of variables}
\setcounter{equation}{0}
The presented algebraic construction are intimately connected
with the method of separation of variables in classical mechanics.
We shall use technique developed by Sklyanin in framework of
$r$-matrix formalism and based on the application of the
Baker-Akhiezer function \cite{skl95}.  Recall, that the
Baker-Akhiezer function $\P (\l)$ is the eigenvector of the Lax
matrix
\bq
L(\l)\P (\l)=z(\l)\P (\l)\,,\quad
\{H,\P\}=\P_t(\l)= A\P(\l)\,,\ll{glax}
\eq
corresponding to the eigenvalue $z(\l)$, which has certain
analyticity properties.
Here $A$ is a second matrix in the Lax representation (\ref{lax}).
Since an eigenvector is defined up to a
scalar factor, to exclude the ambiguity in the definition of $\P (\l)$
one has to fix a normalization of $\P (\l)$ imposing a linear
constraint
\[\sum_{j=1}^N \b_j(\l)\P _j(\l)=1\,.\]
The main purpose of
this Section is to find a correspondence between the outer
automorphisms of infinite-dimensional representations of $sl(2)$ and
the normalization of the Baker-Akhiezer function.

According to the Sklyanin recipe the variables of separation
$(x_j,p_j)$ are defined by the poles of the properly normalized
Baker-Akhiezer function $\P (\l)$ and the corresponding
eigenvalues $z(\l)$ of the Lax matrix \cite{skl95}.  For the
Lax matrix $L(\l)\in \te{sl(2)}^*$ the poles $x_j$ of $\P (\l)$
are zeroes of the following function
\bq
B(\l,\b_1,\b_2)=
\b_1^2(\l) s_+(\l)-\b_2^2(\l) s_-(\l)-2\b_1(\l)\b_2(\l) s_3(\l)\,.
\ll{BBB}
\eq
Introduce functions $\a_j(\l)$ as
\bq
\a_1(\l)=[(\b_1^2-\b_2^2)f](\l)\,,\qquad
\a_2(\l)=i[(\b_1^2+\b_2^2)f](\l)\,,\qquad
\a_3=-2[\b_1\b_2f](\l)\,,\ll{cof}
\eq
where $f(\l)$ is a certain common function of a spectral parameter only.
The coefficients $\a_j(\l)$ lie on the cone (\ref{restr1}) and
define the linear operator $b(\l,\a_j)=B(\l,\b_1,\b_2)$ (\ref{bl})
for the mappings (\ref{map2}) and (\ref{mapmn}).  The second
restriction (\ref{restr2}) on $\a_j(\l)$ guarantees that all zeroes
$x_j$ of $b(\l)$ are mutually commuting $\{x_j,x_k\}=0$.

For the linear $R$-bracket the eigenvalues $z(\l)$ of the Lax matrix
$L(\l)$ corresponding to zeroes $x_j$ are canonically conjugated
momenta $p_j$.  The pairs of variables $(x_j,p_j)$ lie on the
spectral curve
\bq
W(p_j,x_j)=0\,,\qquad
W(z,\l)={\rm det}(z-L(\l))=z^2-\D(\l)\,,
\ll{spc}
\eq
which fits exactly to separated equations \cite{skl95}.

The equations (\ref{glax}) and divisor of poles $\P(\l)$ are covariant
with respect to the mappings (\ref{map2}-\ref{mapmn})
\[L\to L'=L+\D L_{MN}\,,\qquad H\to H'=H+V_{MN}.\]
These mappings could be considered as an
analogous of a standard Darboux transformation.

The mappings (\ref{mapmn}) change the separated equations (\ref{spc})
as
\bq
p_j^2-\D(x_j,I_k)=0\to
p_j^2-\D'(x_j,I'_k)=0\,,
\ll{sepeq}
\eq
where $\D'(\l)$ and new integrals of motion $I'_k$ are given by
(\ref{nint}).  Thus, by taking the single Lax matrix $L(\l)$
associated to some integrable system, which can be integrated
by separation of variables in coordinates $\{x_j,p_j\}$, we
get an infinite set of completely integrable systems
determined by the mappings (\ref{mapmn}), which are separable
in the same variables $\{x_j,p_j\}$.

The linear operator $b(\l)$  (\ref{bl}) is a symmetric function of
its zeroes $\{x_j\}_{j=1}^n$ and dynamical $r$-matrices
(\ref{rsb}) depends of the spectral parameters and only half of the
dynamical variables $\{x_j\}_{j=1}^n$.  Moreover, if normalization
$\b(\l)$ of $\P (\l)$ is given by arbitrary constant numeric vector,
then three functions $\a_j(\l)$ (\ref{cof}) differ by certain numeric
constants $\b_j$
\bq \a_j(\l)=\b_j f(\l)\,,\qquad \b_j\in\bC,
\qquad \sum\b_j^2=0\,.\ll{cnorm} \eq
Then by (\ref{cnorm}) the invariant
polynomial $\D(\l)$ and all the integrals of motion $I_k$ (\ref{ham})
are shifted on items said to be potentials depending of
separated coordinates $x_j$ (\ref{mapi}):
\ben
&&\D'=\D+V_{MN}(x_1,\ldots,x_n)=\D+2b(x_j,\l)\left[
\dfrac{f(\l)}{b(x_j,\l)}\right]_{MN}\,,\nn\\ \ll{poten}\\
&& I'_k=I_k+U_k(x_1,\ldots,x_n)=
I_k+{\rm Res}_{\l=0}\left(\phi_k(\l)
V_{MN}(x_1,\ldots,x_n)\right)\,.\nn
\en
Thus, all integrable systems related to the mappings (\ref{mapmn})
with the constant normalization of the corresponding Baker-Akhiezer
function obey to the Levi-Civita theorem (\ref{gh}) \cite{lc04}.

Integrals of motion $I_j'$ (\ref{intmul}) associated to the
multiplicative automorphism (\ref{mapmug}) could be differ from
original ones on certain functions
$a_{ij}(x_1\ldots,x_n)$ of only half of the dynamical variables
$\{x_j\}_{j=1}^n$
(see (\ref{mapi}) and the St\"ackel systems (\ref{stint})).
Here functions $a_{ij}$ are defined by (\ref{intmul}) for
any concrete representation.

Method of separation of variables for the Lax matrices
$L(\l)\in\te{sl(N)}^*$ was studied \cite{skl95}.  The separated
coordinates are obtained as zeroes of the certain polynomial $B(\l)$
of degree $N(N-1)/2$ in components of $L(\l)$.  The spectral
curve $W(z,\l)={\rm det}(z-L(\l))$ is a nonhyperelliptic algebraic
curve for $N>2$ and the Levi-Civita theorem can not be applied
directly to this case. Still the polynomial $B(\l)$ depending of separated
coordinates  can be used to construct a counterpart of the mappings
(\ref{map2}) and (\ref{mapmn}) for $\te{sl(N)}$.

Consider, for instance, the loop algebra $\te{sl(3)}$.
Let the Lax matrix $L(\l)\in \te{sl(3)}^*$ obeys the standard linear
$R$-bracket (\ref{rpoi}) with rational $r$-matrix
$r_{12}=(\l-\mu)^{-1}\Pi$, where $\Pi$ is the permutation
operator:  $\Pi x\otimes y=y\otimes x,~x,y\in \bC^3$. The
entries $s_{ij}(\l)$ of the Lax matrix $L(\l)$ are constructed
in  variables $s_{ij}~(i,j=1,2,3.~\sum s_{ii}=0)$ with
the following standard Poisson brackets
\[
\{s_{ij},s_{km}\}=s_{im}\delta_{jk}-s_{kj}\delta_{im}\,,\]
which define the natural Lie-Poisson bracket on $\te{sl(3)}^*$.
According to \cite{skl95}, the simplest choice of normalization
$\b(\l)$ of $\P (\l)$ is
\bq \b_1=\b_2=0\,,\qquad \b_3=1\,,\ll{nb} \eq
then the polynomial $B(\l)$ is given by
\bq
B(\l)=s_{32}(\l)u_{13}(\l)-s_{31}(\l)u_{23}(\l)\,,
\ll{b3}
\eq
where $u_{ij}(\l)$ is $(ij)$-cofactor of the determinant of $L(\l)$.
The polynomial $B(\l)$ (\ref{b3}) allows to introduce
the following mapping
\ben
&&s_{ij}\to s'_{ij}=s_{ij}\,,\qquad (ij)\neq (13),(23)\,,\nn\\
&&s_{13}\to s'_{13}=s_{13}+s_{32}f(\l)B^{-1}(\l)\,,
\ll{map3}\\
&&s_{23}\to s'_{23}=s_{23}-s_{31}f(\l)B^{-1}(\l)\,,\nn
\en
as an analog of the mapping (\ref{map2}) with the normalization
(\ref{nb}). This mapping
leaves fixed the spectral invariants $\tau_1$ and $\tau_2$ and
shifts the third invariant polynomial $\tau_3$
\ben
&&\tau_1(\l)={\rm tr} L(\l)=0\to \tau'_1(\l)=\tau_1(\l)=0\nn\\
&&\tau_2(\l)={\rm tr} L^2(\l)\to \tau'_2(\l)=
\tau_2(\l)\,,\nn\\
&&\tau_3(\l)={\rm det} L(\l)\to \tau'_3(\l)=
\tau_3(\l)+f(\l)\,,\nn
\en
The general Lax matrix $L'(\l)$ defined by (\ref{map3}) obeys the
linear $R$-bracket with the dynamical $r$-matrices.
As an example, the Lax representations for the Henon-Heiles
system and a system with quartic potential \cite{bef95} can be
embedded into the proposed scheme by using the more
sophisticated normalization.

So, we can apply the known polynomials $B(\l)$ and method of
separation of variables to construct the analogs of mappings
(\ref{map2}) and (\ref{mapmn}) for the loop algebra
$\te{sl(N)}$. It would be interesting to find the general
counterparts of outer automorphisms (\ref{mapj}) and (\ref{mapmug})
for infinite-dimensional representations of $sl(N)$ and to consider the
inverse problem of a choosing the correct normalization of
Baker-Akhiezer function by using these automorphisms.  A similar, but
more difficult problem arises for the simple Lie algebras other than
$sl(N)$.


\section{Dynamical $r$-matrices and quadratic $R$-bracket}
\setcounter{equation}{0}
Now we consider analogs of the additive
and multiplicative automorphisms of $sl(2)$
(\ref{mapp})-(\ref{mapmu}) for the quadratic $R$-bracket in
classical mechanics.

Introduce the formal particular mapping
\bq
T(u)=
\left(\begin{array}{cc}A&B\\C&D\end{array}\right)\,(u)\to
T'(u)=
\left(\begin{array}{cc}A+fD^{-1}&B\\C&D\end{array}\right)\,(u).
\ll{mapq}
\eq
If the initial matrix $T(u)$ obeys the standard quadratic
$R$-bracket
\bq
\{\on{T}{1}(u), \on{T}{2}(v)\}=
[r(u-v), \on{T}{1}(u)\on{T}{2}(v)\,]\,,\quad
{\rm with}\quad
r=(u-v)^{-1}\Pi\,,\ll{qrb}
\eq
then the image $T'(u)$ of mapping (\ref{mapq}) obeys the
dynamical quadratic $R$-bracket
\bq
\{\on{T'}{1}(u), \on{T'}{2}(v)\}=
[r(u-v), \on{T'}{1}(u)\on{T'}{2}(v)\,]
+\on{T'}{2}(v)s_{21}\on{T'}{1}(u)-
\on{T'}{1}(u)s_{12}\on{T'}{2}(v)\,.\ll{dqb}
\eq
Here dynamical matrices $s_{jk}$ are given by
\ben
&&s_{12}(u,v)=\dfrac{1}{u-v}\left[
\dfrac{f(u)}{D^2(u)}-\dfrac{f(v)}{D^2(v)}\right]\cdot
\sigma_-\otimes\sigma_+\,,\nn\\
\ll{dqr}\\
&&s_{21}(u,v)=\Pi s_{12}(v,u) \Pi\,.\nn
\en
In contrast to the linear case, the mapping (\ref{mapq})
related to additive automorphism (\ref{mapp}) changes the form
of $R$-bracket. However,  if the functions $D(u)$ and $f(u)$
(\ref{mapq}) are independent on spectral parameter $u$, then
the dynamical matrices $s_{jk}$ in (\ref{dqb}) go to
zero \cite{ts94}. Moreover, in this case, we can use the
more general mapping (see (\ref{mapq}))
\[A(u)\to A'(u)=A(u)+g(D)\,,\]
where $g(D)$ is an arbitrary function on entry $D$.
This mapping changes the standard $R$-bracket (\ref{qrb}) to
dynamical bracket (\ref{dqb}), but it preserves the property
of integrability \[\{\tr T'(u),\tr T'(v)\}=0\,.\]
We have to emphasize that additive and multiplicative automorphisms
of $sl(2)$ (\ref{mapp})-(\ref{mapmu}) give rise to
the integrable systems associated to the two root systems $BC_n$ and
$D_n$ \cite{ts94}, respectively.

Consider the special solutions
\bq T_A(u)=\left(\begin{array}{cc}A(u)&B(u)\\
                       C(u)&-A(-u)\end{array}\right)\,,
\eq
of the classical reflection equation
\ben
\left\{\on{T}{1}(u), \on{T}{2}(v)\right\}&=&
\left[r(u-v), \on{T}{1}(u)\on{T}{2}(v)\right]+\ll{repoi}\\
&+&\on{T}{1}(u)r(u+v)\on{T}{2}(v)-
\on{T}{2}(v)r(u+v)\on{T}{1}(u). \nn
\en
with the rational $r=(u-v)^{-1}\Pi$ $r$-matrix.
An application of the additive and multiplicative automorphisms
of $sl(2)$ (\ref{mapp})-(\ref{mapmu}) is more tricky in the
quadratic case.

If the  entry $B(u)$ is independent of spectral parameter
$u$, i.e. $\partial B/\partial u=0$, we can construct  the new
solution of the reflection equation (\ref{repoi})
\bq
T_{C}=T_A+
\left(\begin{array}{cc}0 &0\\
\gamma B^{-1}&0 \end{array}\right)\,,\qquad
\g\in \bR\,.\ll{tc}
\eq
Assuming in addition that entry $A(u)$ is a linear function
of $u$, i.e. $A(u)=ua_1+a_2$, let us introduce the second
bounary matrix
\ben
T_{BC}&&=T_C+
\left(\begin{array}{cc}\dfrac{\a}{u}+\b &0\\
\left[\dfrac{\a}{u}(A(u)-A(-u))+\b(A(u)+A(-u))\right]B^{-1}
&\dfrac{\a}{u}-\b \end{array}\right)\,,\nn\\
\ll{tbc}\\ \nn\\
&&\qquad\a,\b,\in \bR\nn
\en
which is new solution of reflection equation (\ref{repoi}).
These solution $T_C$ (\ref{tc}) and $T_{BC}$ (\ref{tbc})
could be associated with the additive automorphism
(\ref{mapp}).

If the central element $\D=\det T_A(u)=0$ is
equal to zero and $\partial B/\partial u=0$
the third boundary matrix
\bq
T_{D}=
\left(\begin{array}{cc}A(u)-A(-u)B^{-1}&B\cdot(1-B^{-1})^2\\
C(u)&-A(-u)+A(u)B^{-1}\end{array}\right)\,. \ll{td}
\eq
is a solution of reflection equation (\ref{repoi}).  This
solution could be associated with the multiplicative automorphism
(\ref{mapmu}).

As an example, we consider the Toda lattices.  Let the initial
boundary matrix $T_A(u)$ is given by
\bq T_A(u)=\left(\begin{array}{cc}(u-p)\exp(q)&\exp(2q)\\
	   u^2-p^2&(u+p)\exp(q)\end{array}\right)\,,
\eq
where $(q,p)$ is a pair of canonically conjugate variables.
According to \cite{ts94} matrices $T_A$, $T_{BC}$ and $T_D$
correspond to the Toda lattices associated
with the Lie algebras of $A_n$, $B_n\,(\b=\g=0)$,
$C_n\,(\a=\b=0)$ and $D_n$ series, respectively.
Among the hamiltonians, in comparison with (\ref{mapi}), there are
\ben
&H_A&=\dfrac12\sum_{j=1}^n
p_j^2+\sum_{j=1}^{n-1}\exp(x_{j+1}-x_j)\,,\nn\\
\nn\\
&H_{BC}&=H_A+\g\exp(-2x_1)+(2\a+2\b p_1)\exp(-x_1)\,,\nn\\
&H_D&=H_A+\exp(-x_1-x_2)\,.\nn
\en
In classical and quantum mechanics the boundary matrices
(\ref{tbc}) and (\ref{td}) have been used for the relativistic
Toda lattices and for the Heisenberg $XXX$ and $XXZ$ models in
\cite{ts94}.

The more complicated dynamical quadratic $R$-bracket has been
introduced in \cite{ts95} for the Neumann top, Kowalewski top
and Toda lattice associated to the Lie algebra $G_2$.

\section{Conclusions}
\setcounter{equation}{0}
The outer automorphisms of infinite-dimensional representation
of $sl(2)$ give possibility to construct new  Lax matrices.
The corresponding linear and quadratic $R$-brackets include the
dynamical $r$-matrices, which obey the dynamical Yang-Baxter
equations.

For the loop algebras as a second step we applied the certain
projection of  general Lax matrix onto the low-dimensional
subspaces, which preserve the $R$-bracket. Thus, the set of the
Lax matrices associated to the different integrable system can
be obtained.

The similar dynamical deformations of quadratic $R$-bracket
has been applied for the construction of Lax matrix for
integrable systems associated to the root systems $BC_n$ and $D_n$.

Among the known and possible examples there are:  a wide class
of the St{\"a}ckel systems; the integrable extensions of the
classical tops - Euler top, Manakov and Steklov tops, Lagrange
and Goryachev-Chaplygin top; generalizations of the Heisenberg
and Gaudin magnets; the Toda and the Calogero-Moser systems.

The following problems, however, remain open.
We have not an exhaustive description of outer automorphisms of
infinite-dimensional representations of simple Lie algebras and of
the all admissible low-dimensional submanifolds, which allow a
well-defined restriction to the corresponding linear classical and
quantum $R$-brackets.

\section{Acknowledgments}
\setcounter{equation}{0}
We would like to thank I.V. Komarov for reading the manuscript and
making valuable comments. This research has been partially supported
by RFBR grant 96-0100537.


\end{document}